\documentclass[aps,preprint]{revtex4}

\usepackage{graphics}
\usepackage{amssymb}
\usepackage[]{amsmath, amsfonts, bm}

\begin{document}
%
\title{Fourth-rank tensors of $[[V^{2}]^{2}]$ type and elastic material constants for all systems of 2D crystals}
\author{Cz. Jasiukiewicz}
\author{T. Paszkiewicz}
\email{tapasz@prz.edu.pl} 
\author{S. Wolski}
\begin{abstract}
Fourth-rank tensors $[[V^{2}]^{2}]$ (Voigt's) type, that embody the elastic properties of crystalline anisotropic substances, were constructed for all 2D crystal systems. Using them we obtained explicit expressions for inverse of Young's modulus $E(\textbf{n})$, inverse of shear modulus  $G(\textbf{m},\textbf{n})$ and Poisson's ratio $\nu(\textbf{m},\textbf{n})$, which depend on components of the elastic compliances tensor $\textbf{S}$, on direction cosines of vectors $\textbf{n}$ of uniaxial load and the vector $\textbf{m}$ of lateral strain with crystalline symmetry axes. All 2D crystal systems are considered. Such representation yields decomposition of the above elastic characteristics to isotropic and anisotropic parts. Expressions for Poisson's coefficient are well suited for studying the property of auxeticity and anisotropy 2D crystals. 
Phase velocities are calculated for all 2D crystal classes.
\end{abstract}

\affiliation{Chair
of Physics, Rzesz\'{o}w University of Technology, Al. Powsta\'{n}c\'{o}w. Warszawy 6,
PL-35-959 Rzesz\'{o}w Poland} 
\maketitle

\section{Introduction} 
\label{introd}
In our previous paper we derived explicit expressions for inverse of Young's modulus $E(\textbf{n})$, inverse of shear modulus  $G(\textbf{m},\textbf{n})$ and Poisson's ratio $\nu(\textbf{m},\textbf{n})$ for 3D crystal systems of high and middle symmetry \cite{pawol}. Our derivation yields expressions which do not depend on the choice of Cartesian coordinate system. These expressions depend on components of the compliances tensor $\textbf{S}$, and on direction cosines of vectors $\textbf{n}$ of uniaxial load and the vector $\textbf{m}$ of lateral strain with crystalline symmetry axes. To obtain such formulas, we used fourth-rank tensorial bases introduced by Walpole \cite{walpole}. He considered such tensorial bases for all thirty-two 3D crystal classes. The fourth-rank tensors of Voigt's type for 2D crystalline solids do not seem to have been submitted previously, therefore, we construct and use them to calculate the above mentioned characteristics of elastic media belonging to all four 2D crystal systems \cite{musgrave}. 

Auxetic elastic materials exhibits unusual property of becoming wider when stretched and thinner when compressed, i.e. they exhibit a negative Poisson ratio. Love \cite{love} presents a single example of cubic single crystal pyrite as having a Poisson's ratio of -0.14. Lakes described the synthesis of an actual auxetic material and proposed a simple mechanism underlying the negative Poisson's ratio \cite{lakes}. Alderson and Evans fabricated microporous polyethylene having a negative Poisson's ratio \cite{alderson-evans}. Data were compiled from the literature by Baughman et al. and were analyzed to show negative Poisson's ratio to occur for stress in an oblique direction upon single crystals of cubic metals \cite{baughman'98}.

The unusual properties of auxetics imparts many beneficial effects on the material's macroscopic properties that make auxetics superior to conventional materials in many commercial applications. To explain microscopic mechanisms underlying the property of auxeticity, various models are proposed. Two-dimensional microscopic crystalline models make an interesting class of auxetics (cf. e.g. \cite{KWW}-\cite{tret-woj}). Each model of such kind belongs to one of crystal systems, therefore its auxetic properties are characterized by Poisson's ratio suitable for this system.

For completeness, in the Appendix we consider the Christoffel tensor for all 2D crystal Laue groups in the Appendix to this paper. 
\section{Bases of fourth-rank tensors for 2D crystalline media}
Repetition of a latin suffix in a product of tensors or in a single tensor implies summation with respect to that suffix over values 1,2. If $\textbf{u}$, $\textbf{v}$ are two mutually perpendicular vectors, then 	
\begin{equation}
u_{i}u_{i}=1,\; u_{i}v_{i}=0,\;u_{i}u_{j}+v_{i}v_{j}=\delta_{ij},	
\label{eq:vector-ident}
\end{equation}
the last of these relations can be written in a dyadic form $\left(\textbf{u}\otimes\textbf{v}+\textbf{v}\otimes\textbf{u}\right)=I_{2}$. 

The stiffness tensor $C$ and the compliances tensor $S$ are mutually inverse, i.e. 
\begin{equation}
	SC=CS=I_{4}.
	\label{eq:CS}
\end{equation}
The product $AB$ of two fourth-rank tensors $A$ and $B$ has components $(AB)_{jkl}=A_{ijrs}B_{rskl}$. 

The fourth rank identity tensor $I_{4}$ has components 
	\[
\left(I_{4}\right)_{ijkl}=\left(\delta_{ik}\delta_{jl}+\delta_{il}\delta_{jk} \right)/2.	
\]

\subsection{Hexagonal (isotropic) system}
\label{hexagonal}
Consider crystal belonging to the hexagonal (isotropic) system. Two tensors $J$ and $K$ 
	\[
	J_{ijkl}=\delta_{ij}\delta_{kl}/2,\;
	K_{ijkl}=\left(\delta_{ik}\delta_{jl}+\delta_{il}\delta_{jk}-\delta_{ij}\delta_{kl}\right)/2,
\]
with the multiplication table $J^{2}=J$, $K^{2}=K$, $JK=KJ=0$ bring about the decomposition $J+K=I_{4}$. The general fourth-rank tensor of complete Voigt's symmetry (i.e. of $[[V^{2}]^{2}]$ type) is constructed as 
\begin{equation}
	A=a_{J}J+a_{K}K,
\label{eq:A-decomp}
\end{equation}
where 
	\begin{equation}
	a_{J}=A_{iijj}/2,\; a_{K}=\left(A_{ijij}-a_{J} \right)/2.
\label{eq:A-coeff}
\end{equation}
Applying Eqs. (\ref{eq:A-decomp}) and (\ref{eq:A-coeff}) to $C$ and $S$ tensors, we obtain 
\begin{eqnarray}
	c_{J}=C_{11}+C_{12}, \; c_{K}=C_{11}-C_{12},\nonumber \\
	s_{J}=S_{11}+S_{12}, \; s_{K}=S_{11}-S_{12},
\label{eq:C-S-exp-coeff}	
\end{eqnarray}
where $C_{11}\equiv C_{1111},\, C_{12}\equiv C_{1122}$. Similarly, $S_{11}\equiv S_{1111},\, S_{12}\equiv S_{1122}$.

An isotropic 2D medium is mechanically stable if
\begin{equation}
	s_{J}=S_{11}+S_{12}>0,\; s_{K}=S_{11}-S_{12}>0, 
	\label{eq:stabil-isotrop}
\end{equation}
i.e. $S_{11}>|S_{12}|$. From inequalities (\ref{eq:stabil-isotrop}) it is easily shown that $S_{11}>0$. Eigenvalues $c_{J},\;c_{K}$ and $C_{11}$, $C_{12}$ obey analogous inequalities. 

Using Eqs. (\ref{eq:CS}) and multiplication table for $J$ and $K$, we express components of $S$ by components of $C$
	\begin{equation}
	S_{11}=\frac{C_{11}}{C_{11}^{2}-C_{12}^{2}},\; S_{12}=-\frac{C_{12}}{C_{11}^{2}-C_{12}^{2}}. 
\label{eq:S-versus-C}
\end{equation}

\subsection{Quadratic system}
\label{quadratic}
Two crystallographic directions of the quadratic system, $\textbf{a}$ and $\textbf{b}$, are mutually perpendicular, hence they obey the relations (\ref{eq:vector-ident}). Now the tensors $K$ and $I_{4}$ decompose
	\begin{equation}
	K=L+M, \; I_{4}=J+L+M.
\nonumber
\end{equation}
The tensors $L=\left(\textbf{u}\otimes\textbf{u}-\textbf{v}\otimes\textbf{v}\right)\otimes\left(\textbf{u}\otimes\textbf{u}-\textbf{v}\otimes\textbf{v}\right)/2$ and $M=\left(\textbf{u}\otimes\textbf{v}+\textbf{v}\otimes\textbf{u}\right)\otimes\left(\textbf{u}\otimes\textbf{v}+\textbf{v}\otimes\textbf{u}\right)/2$ are idempotent and have components 
\begin{eqnarray}
L_{ijkl}=\left(u_{i}u_{j}-v_{i}v_{j}\right)\left(u_{k}u_{l}-v_{k}v_{l}\right)/2,\\
M_{ijkl}=\left(u_{i}v_{j}+v_{i}u_{j}\right)\left(u_{k}v_{l}+v_{k}u_{l}\right)/2.
\label{eq:L-M}	
\end{eqnarray}
The tensors $J$, $L$, and $M$ fulfill the multiplication table 
\begin{eqnarray}
	L^{2}=L,\; M^{2}=M,\; JL=LJ=0,\; JM=MJ=0\; LM=ML=0.
	\label{eq:quadr-mult-table}
\end{eqnarray}

The general fourth-rank tensor of the $[[V^{2}]^{2}]$ type is composed as 
\begin{equation}
	A=a_{J}J+a_{L}L+a_{M}M, 
\label{eq:voigt-q-decomp}
\end{equation}
where 
\begin{eqnarray}
	a_{L}=\left(u_{i}u_{j}-v_{i}v_{j}\right)A_{ijkl}\left(u_{k}u_{l}-v_{k}v_{l}\right)/2,\nonumber\\ 
  a_{M}=\left(u_{i}v_{j}+v_{i}u_{j}\right)A_{ijkl}\left(u_{k}v_{l}+v_{k}u_{l}\right)/2.
\label{eq:quadr-coeff}
\end{eqnarray}

Using Eqs. (\ref{eq:voigt-q-decomp}) and (\ref{eq:quadr-coeff}) for $C$ and $S$ tensors, we obtain  in the reference axes coincident with the crystal reference axes (CRA for short)    
\begin{equation}
	c_{L}=\left(C_{11}-C_{12}\right),\; c_{M}=2C_{1212}, \; s_{L}=\left(S_{11}-S_{12}\right),\; s_{M}=2S_{1212}. 
	\label{eq:quadr-comp}
\end{equation}
A quadratic medium is mechanically stable if $c_{\sigma}>0$ or $s_{\sigma}>0$ ($\sigma=J,L,M$). In terms of $S_{IJ}\; I,J=1,2,6$
\begin{equation}
	S_{11}>0,\; S_{66}>0, 
\label{quadratic-stab}
\end{equation}
and conditions (\ref{eq:stabil-isotrop}) are fulfilled, i.e. $S_{11}>|S_{12}|$. 

The components of $S$ are related to components of $C$. Eqs. (\ref{eq:S-versus-C}) are fulfilled and, since $C_{66} \equiv C_{1212}$ and $S_{66} \equiv 4S_{1212}$ \cite{musgrave},
\begin{equation}
	S_{66}=1/C_{66}.
	\label{quadr-SversusC}
\end{equation}

When $a_{L}=a_{M}$, the general symmetric fourth rank-tensor $A$ reduces to the corresponding tensor for the isotropic 2D system. For $A=C$, one obtains the familiar relation $C_{66}=\left(C_{11}-C_{12}\right)/2$ and $S_{66}=2\left(S_{11}-S_{12}\right)$ for $A=S$. 
\subsection{Rectangular system }
\label{sc:rectang}
As the crystallographic directions of all classes of the rectangular system are mutually perpendicular, the conditions (\ref{eq:vector-ident}) are satisfied by the unit vectors $\textbf{u}$, $\textbf{v}$ that define these directions. The tensor $M$ is retained intact. The four new tensors are introduced 
\begin{eqnarray}
E_{11}=(\textbf{u}\otimes\textbf{u})\otimes(\textbf{u}\otimes\textbf{u}), \; E_{12}=(\textbf{u}\otimes\textbf{u})\otimes(\textbf{v}\otimes\textbf{v}),\nonumber \\
E_{21}=(\textbf{v}\otimes\textbf{v})\otimes(\textbf{u}\otimes\textbf{u}), \; E_{22}=(\textbf{v}\otimes\textbf{v})\otimes(\textbf{v}\otimes\textbf{v}).
\label{eq:rectang-base-tens}  
\end{eqnarray}

The tensors $M$, $E_{pq}$, ($p,q=1,2$) form the complete set whose multiplication table is 
\begin{table}
\begin{tabular}
[c]{c|ccccc}
& $E_{11}$ & $E_{22}$ & $E_{12}$ & $E_{21}$ & $M$\\ \hline
$E_{11}$ & $E_{11}$ & 0 & $E_{12}$ & 0 & 0\\
$E_{22}$ & 0 & $E_{22}$ & 0 & $E_{21}$ & 0\\
$E_{12}$ & 0 & $E_{12}$ & 0 & $E_{11}$ & 0\\
$E_{21}$ & $E_{21}$ & 0 & $E_{22}$ & 0 & 0\\
$M$ & 0 & 0 & 0 & 0 & $M$
\end{tabular}
\end{table}
The general fourth-rank tensor of the $[[V^{2}]^{2}]$ type is constructed by the linear combination 
\begin{equation}
	A=a_{11}E_{11}+a_{22}E_{22}+a_{s}E_{s}+a_{M}M,
	\label{eq:rect-decomp}
\end{equation}
where $E_{s}=(E_{12}+E_{21})$ and
\begin{eqnarray}
	a_{11}=u_{i}u_{j}A_{ijkl}u_{k}u_{l},\; 	a_{22}=v_{i}v_{j}A_{ijkl}v_{k}v_{l}, \nonumber \\
	a_{12}=u_{i}u_{j}A_{ijkl}v_{k}v_{l}=a_{21}\equiv a_{s},\nonumber \\ 
	a_{M}=\left(u_{j}v_{j}+v_{i}u_{j}\right)A_{ijkl}\left(u_{k}v_{l}+v_{k}u_{l}\right)/2. 
\label{eq:rectang-gen-coeff}
\end{eqnarray}

In CRA, by combining Eqs. (\ref{eq:rect-decomp}) and (\ref{eq:rectang-gen-coeff}) for $C$ and $S$ tensors, we can evaluate $c_{\alpha \beta}\; (\alpha, \beta =1,2,M)$ as follows 
\begin{eqnarray}
	c_{11}=C_{11},\; c_{22}=C_{22},\; c_{12}=c_{21}=C_{12},\; c_{M}=2C_{1212}=2C_{66}, \nonumber \\
	s_{11}=S_{11},\; s_{22}=S_{22},\; s_{12}=s_{21}=S_{12},\; s_{M}=2S_{1212}=S_{66}/2. 
	\label{eq:rectang-comp}
\end{eqnarray}

It follows from Eqs. (\ref{eq:rect-decomp}) and (\ref{eq:rectang-gen-coeff}) that
\begin{eqnarray}
	J=\left(E_{11}+E_{22}+E_{s}\right)/2,\; K=\left(E_{11}+E_{22}-E_{s}\right)/2+M,\nonumber \\
	L=\left(E_{11}+E_{22}-E_{s}\right)/2,\; I_{4}=E_{11}+E_{22}+E_{M}. 
	\label{eq:rect-IJKdecomp}
\end{eqnarray}

In the CRA, relations connecting compliances and stiffnesses of the rectangular system read
\begin{eqnarray}
	S_{11}=\frac{C_{22}}{C_{11}C_{22}-C_{12}^{2}}, \; S_{22}=\frac{C_{11}}{C_{11}C_{22}-C_{12}^{2}},\nonumber \\ S_{12}=-\frac{C_{12}}{C_{11}C_{22}-C_{12}^{2}},\; S_{66}=C_{66}^{-1}.
\label{eq:rect-SversusC}
\end{eqnarray}

A rectangular elastic medium is mechanically stable if inequalities (\ref{quadratic-stab}) are fulfilled and 
\begin{equation}
	S_{22}>0,\;\; S_{11}S_{22}>S_{12}^{2}.
	\label{rect-stab}
\end{equation}
\subsection{Oblique crystal system}
\label{parallelogram}
Two unit vectors $\textbf{a}$ and $\textbf{b}$ make an arbitrary angle ($\neq\pi/2$) with each other in the plane that they define. It is convenient to introduce two further unit vectors $\textbf{u}$, $\textbf{v}$ which are mutually perpendicular as independent combinations of $\textbf{a}$ and $\textbf{b}$, e.g. $\textbf{a}=\textbf{u}$, $\textbf{v}=\left[\textbf{b}-(\textbf{a}\textbf{b})\textbf{a}\right]/\left[1-(\textbf{a}\textbf{b})^{2}\right]^{1/2}$ \cite{walpole}. 

The tensors $E_{\alpha\alpha}\:(\alpha =1,2)$  and $M\equiv E_{33}$ are retained intact. Four new tensors are introduced 
\begin{eqnarray}
E_{13}=(\mathbf{u}\otimes\mathbf{u})\otimes(\mathbf{u}\otimes\mathbf{v}+\mathbf{v}\otimes\mathbf{u})/\sqrt{2},
\nonumber \\ 
E_{31}=(\mathbf{u}\otimes\mathbf{v}+\mathbf{v}\otimes\mathbf{u})\otimes(\mathbf{u}\otimes\mathbf{u})/\sqrt{2}, 
\nonumber \\
E_{23}=(\mathbf{v}\otimes\mathbf{v})\otimes(\mathbf{u}\otimes\mathbf{v}+\mathbf{v}\otimes\mathbf{u})/\sqrt{2},
\nonumber \\E_{32}=(\mathbf{u}\otimes\mathbf{v}+\mathbf{v}\otimes\mathbf{u})\otimes(\mathbf{v}\otimes\mathbf{v})/\sqrt{2}. \label{eq:parallel-tensors}
\end{eqnarray}
The multiplication table 

\begin{table}
\begin{tabular}
[c]{c|ccccccccc}
& $E_{11}$ & $E_{22}$ & $E_{33}$ & $E_{12}$ & $E_{21}$ & $E_{13}$ &
$E_{31}$ & $E_{23}$ & $E_{32}$\\ \hline
$E_{11}$ & $E_{11}$ & 0 & 0 & $E_{12}$ & 0 & $E_{13}$ & 0 & 0 & 0\\
$E_{22}$ & 0 & $E_{22}$ & 0 & 0 & $E_{21}$ & 0 & 0 & $E_{23}$ & 0\\
$E_{33}$ & 0 & 0 & $E_{33}$ & 0 & 0 & 0 & $E_{31}$ & 0 & $E_{32}$\\
$E_{12}$ & 0 & $E_{12}$ & 0 & 0 & $E_{11}$ & 0 & 0 & $E_{13}$ & 0\\
$E_{21}$ & $E_{21}$ & 0 & 0 & $E_{22}$ & 0 & $E_{23}$ & 0 & 0 & 0\\
$E_{13}$ & 0 & 0 & 0 & $E_{13}$ & 0 & 0 & 0 & $E_{11}$ & $E_{12}$\\
$E_{31}$ & $E_{31}$ & 0 & 0 & $E_{32}$ & 0 & $E_{33}$ & 0 & 0 & 0\\
$E_{23}$ & 0 & 0 & $E_{23}$ & 0 & 0 & 0 & $E_{21}$ & 0 & $E_{22}$\\
$E_{32}$ & 0 & $E_{32}$ & 0 & 0 & $E_{31}$ & 0 & 0 & $E_{33}$ & 0
\label{table-paralleogr}
\end{tabular}
\end{table}
can be now compiled. 
	
The pairs tensors $E_{12}$ and $E_{21}$, $E_{13}$ and $E_{31}$, $E_{23}$ and $E_{32}$ have to be made symmetric if the general \emph{symmetric} tensor is required, hence we introduce three symmetric tensors, namely
\begin{equation}
	E^{(s)}_{\alpha\beta}=\left(E_{\alpha\beta}+E_{\beta\alpha}\right) \; (\alpha\neq \beta,\, \alpha,\beta=1,2,3)
\label{symm-tensors}
\end{equation}
Using tensors $E_{\alpha\alpha}\: $ and $E^{(s)}_{\alpha\beta}\: (\alpha\neq \beta)$, ($\alpha, \beta=1,2,3$), the general fourth-rank tensor of $[[V^{2}]^{2}]$ can be constructed for the crystal \emph{2} system
\begin{equation}
	A_{ijkl}=\sum_{\alpha,\beta=1}^{3}a_{\alpha \beta}E^{(\alpha \beta)}_{ijkl},
	\label{eq:parallell-tensor-expansion}
\end{equation}
where $a_{\alpha \beta}=a_{\beta\alpha}$ for $\alpha\neq\beta$ and 
\begin{equation}
	a_{\alpha\beta}=A_{ijkl}E_{ijkl}^{(\alpha\beta)}. 
\label{eq:parallel-exp-coef}
\end{equation}
When $a_{13}=a_{23}=0$, Eq. (\ref{eq:parallel-exp-coef}) reduces to Eq. (\ref{eq:rect-decomp}). 
 
The formulas (\ref{eq:parallell-tensor-expansion}) and (\ref{eq:parallell-tensor-expansion}) bring about decomposition
\begin{eqnarray}
	J=\left[E_{11}+E_{22}+E_{12}^{(s)}\right]/2,\; K=\left[E_{11}+E_{22}-E_{12}^{(s)}\right]/2+E_{33},\nonumber \\
	\; L=\left[E_{11}+E_{22}-E_{12}^{(s)}\right]/2,\; I_{4}=E_{11}+E_{22}+E_{33}.
\label{basic-tensors-expansion}
\end{eqnarray}

Invoking the definitions of the tensors $E_{\alpha \beta}$ and Eq. (\ref{eq:parallel-exp-coef}), we can calculate the elements of the matrix representing the tensors $C$ and $S$ in the CRA
\begin{eqnarray}
	c_{11}=C_{11},\; c_{22}=C_{22},\; c_{33}=2C_{66},\; c_{12}=c_{21}=C_{12},\nonumber \\
	c_{13}=c_{31}=\sqrt{2}C_{1112}\equiv \sqrt{2}C_{16},\; c_{23}=c_{32}=\sqrt{2}C_{2212}\equiv \sqrt{2}C_{26}, 
	\label{eq:parallel-C-coeff}
\end{eqnarray}
\begin{eqnarray}
	s_{11}=S_{11},\; s_{22}=S_{22},\; s_{33}=S_{66}/2,\; s_{12}=s_{21}=S_{12},\nonumber \\
	s_{13}\equiv s_{31}=\sqrt{2}S_{1112}\equiv S_{16}/\sqrt{2},\;s_{23}=s_{32}=\sqrt{2}S_{2212}\equiv S_{26}/\sqrt{2}. 
	\label{eq:parallel-S-coeff}
\end{eqnarray}

Using Eq. (\ref{eq:CS}), we obtain the set of six linear equations for $s_{\alpha \beta}$. Solving them we obtain expressions relating the elements of the matrices representing tensors $S$ and $C$ in the CRA 
\begin{eqnarray}
	S_{11}=\frac{C_{22}C_{66}-C_{26}^{2}}{D},\, S_{12}=\frac{C_{12}C_{66}-C_{16}C_{26}}{D},\nonumber \\ 
	S_{22}=\frac{C_{11}C_{66}-C_{16}^2}{D},\;S_{66}=\frac{C_{11}C_{22}-C_{12}^{2}}{D},\nonumber \\ 	  
	S_{16}=\frac{C_{22}C_{16}-C_{12}C_{26}}{D},\,S_{26}=\frac{C_{11}C_{26}-C_{12}C_{16}}{D},
	\label{eq:paralell-s}
\end{eqnarray}
where $D=\left(C_{11}C_{22}C_{33}+2C_{12}C_{26}C_{16}-C_{22}C_{16}^{2}-C_{11}C_{26}^{2}-C_{12}^{2}C_{66}\right)$.

The stability conditions for oblique crystalline media read
\begin{eqnarray}
	S_{11}>0,\; S_{22}>0,\; S_{66}>0,\; S_{11}S_{22}>S_{12}^{2},\nonumber \\
	S_{11}S_{26}^{2}-2S_{12}S_{26}S_{16}+S_{22}S_{16}^{2}<S_{66}\left(S_{11}S_{22}-S_{12}^{2}\right).
	\label{oblique-stabil}
\end{eqnarray}
\section{Elastic material constants for 2D crystalline media}
\label{sc:definitions}  
Consider any two specified orthogonal unit vectors $\textbf{n}$ and $\textbf{m}$ and four elastic material constants of a considered body, namely the bulk modulus $\kappa$, Young's and shear modules $E(\textbf{n})$, and $G(\textbf{m},\textbf{n})$ respectively, and Poisson's ratio $\nu(\textbf{m},\textbf{n})$. They are defined as \cite{sir-sha,rychl}
\begin{eqnarray}
\kappa^{-1}=I_{2}\cdot S\cdot I_{2},\\
\label{eq:bulk}
\frac{1}{E(\textbf{n})}=(\textbf{n}\otimes\textbf{n})\cdot S\cdot(\textbf{n}\otimes\textbf{n})=n_{i}n_{j}S_{ijkl}n_{k}n_{l},\\
\label{eq:young}
\frac{1}{4G(\textbf{m},\textbf{n})}=(\textbf{m}\otimes\textbf{n})\cdot S\cdot(\textbf{m}\otimes\textbf{n})=m_{i}n_{j}S_{ijkl}m_{k}n_{l},\\ 
\label{eq:shear}
-\frac{\nu(\textbf{m},\textbf{n})}{E(\textbf{n})}=(\textbf{m}\otimes\textbf{m})\cdot S\cdot(\textbf{n}\otimes\textbf{n})=m_{i}m_{j}S_{ijkl}n_{k}n_{l}.
\label{eq:poisson}
\end{eqnarray}

Since $\kappa$, $E^{-1}(\textbf{n})$ and $[4G(\textbf{m},\textbf{n})]^{-1}$ are defined by quadratic forms of positive definite tensor, they are positive, i.e. $\kappa>0$, and $E^{-1}(\textbf{n})>0$, $[4G(\textbf{m},\textbf{n})]^{-1}>0$ for all mutually perpendicular vectors $\textbf{n}$ and $\textbf{m}$. If $\nu(\textbf{m},\textbf{n})$ is negative an elastic medium is said to be an auxetic. 
\subsection{Isotropic 2D media}
\label{sc:mechan-charcter-isotropic}
Using definitions (\ref{eq:poisson})-(\ref{eq:shear}), we obtain
\begin{eqnarray}
E^{-1}=s_{11}=S_{11}=\frac{C_{11}}{C_{11}^{2}-C_{12}^{2}},\;\;\frac{\nu}{E}=-s_{12}=-S_{12}=\frac{C_{12}}{C_{11}^{2}-C_{12}^{2}}, \label{nu-isotropic}\\ G^{-1}=2s_{K}=\frac{2}{\left(C_{11}-C_{12}\right)},\; \kappa^{-1}_{is}=2s_{J}=2\left(S_{11}+S_{12}\right)=\frac{2}{C_{11}+C_{12}}.
\label{G-isotropic}
\end{eqnarray}
All these elastic material constants  are isotropic, and when conditions (\ref{eq:stabil-isotrop}) are fulfilled, $\kappa$, $E$ and $G$ are positive. For all remaining crystal classes the bulk modulus is equal to $\kappa_{is}$.
\subsection{Quadratic 2D media}
\label{mechan-character-quadratic} 	 
In the case of 2D quadratic media $E(\textbf{n})$, $G(\textbf{m},\textbf{n})$ and $\nu(\textbf{m},\textbf{n})$ are anisotropic, as they depend on the directional cosines $(\textbf{n}\textbf{u})$,  $(\textbf{n}\textbf{v})$ and $(\textbf{m}\textbf{u})$,  $(\textbf{m}\textbf{v})$
\begin{eqnarray}
E^{-1}(\textbf{n})=	s_{J}/2+s_{L}\left[(\textbf{n}\textbf{u})^{2}- (\textbf{n}\textbf{v})^{2}\right]^{2}/2+2s_{M}(\textbf{n}\textbf{u})^{2}(\textbf{n}\textbf{v})^{2},\nonumber \\
\left[4G(\textbf{m},\textbf{n})\right]^{-1}=s_{L}\left[(\textbf{m}\textbf{u})(\textbf{n}\textbf{u})- (\textbf{m}\textbf{v})(\textbf{n}\textbf{v})\right]^{2}/2\nonumber \\+s_{M}\left[(\textbf{m}\textbf{u})(\textbf{n}\textbf{v})+ (\textbf{m}\textbf{v})(\textbf{n}\textbf{u})\right]^{2}/2,\nonumber\\
-\nu(\textbf{m},\textbf{n})/E(\textbf{n})=s_{J}/2+s_{L}\left[(\textbf{m}\textbf{u})^{2}- (\textbf{m}\textbf{v})^{2} \right]\left[(\textbf{n}\textbf{u})^{2}- (\textbf{n}\textbf{v})^{2} \right]/2\nonumber\\ +2s_{M}(\textbf{m}\textbf{u})(\textbf{m}\textbf{v})(\textbf{n}\textbf{u})(\textbf{n}\textbf{v}).
\label{nu-quadratic} 
\end{eqnarray}

For mechanically stable media, both $E(\textbf{n})$ and $G(\textbf{n},\textbf{m})$ are positive for all mutually perpendicular vectors $\textbf{n}$ and $\textbf{m}$. 
\subsection{Rectangular 2D media}
\label{mechan-character-rectang}
In the case of rectangular 2D crystalline media, expressions for elastic material constants are more complex than for isotropic and quadratic media

\begin{eqnarray}
E^{-1}(\textbf{n})=s_{11}(\textbf{n}\textbf{u})^{4}+s_{22}(\textbf{n}\textbf{v})^{4}+	2\left(s_{12}+s_{M}\right)(\textbf{n}\textbf{u})^{2}(\textbf{n}\textbf{v})^{2}, \nonumber \\
-\nu(\textbf{m},\textbf{n})/E(\textbf{n})=s_{11}(\textbf{m}\textbf{u})^{2}(\textbf{n}\textbf{u})^{2}+s_{22}(\textbf{m}\textbf{v})^{2}(\textbf{n}\textbf{v})^{2}\nonumber \\ +s_{12}\left[(\textbf{m}\textbf{u})^{2}(\textbf{n}\textbf{v})^{2}+(\textbf{m}\textbf{v})^{2}(\textbf{n}\textbf{u})^{2}\right]+2s_{M}(\textbf{m}\textbf{u})(\textbf{m}\textbf{v})(\textbf{n}\textbf{u})(\textbf{n}\textbf{v}),\label{nu-rect} \\
\left[4G(\textbf{m},\textbf{n})\right]^{-1}=s_{11}(\textbf{m}\textbf{u})^{2}(\textbf{n}\textbf{u})^{2}+s_{22}(\textbf{m}\textbf{v})^{2}(\textbf{n}\textbf{v})^{2}\nonumber \\+2s_{12}(\textbf{m}\textbf{u})(\textbf{n}\textbf{u})(\textbf{m}\textbf{v})(\textbf{n}\textbf{u})
+s_{M}\left[(\textbf{m}\textbf{u})(\textbf{n}\textbf{v})+(\textbf{m}\textbf{v})(\textbf{n}\textbf{u})\right]^{2}/2.\nonumber
\end{eqnarray}
\subsection{Oblique 2D media}
\label{sc:mechan-character-oblique}
In the case of oblique 2D crystalline media, we obtain 
\begin{eqnarray}
E^{-1}(\textbf{n})=s_{11}(\textbf{n}\textbf{u})^{4}+s_{22}(\textbf{n}\textbf{v})^{4}+2\left(s_{33}+s_{12}\right)(\textbf{n}\textbf{u})^{2}(\textbf{n}\textbf{v})^{2}\nonumber \\
	+2\sqrt{2}s_{13}(\textbf{n}\textbf{u})^{3}(\textbf{n}\textbf{v})+2\sqrt{2}s_{23}(\textbf{n}\textbf{u})(\textbf{n}\textbf{v})^{3},\nonumber \\
-\nu(\textbf{m},\textbf{n})/E(\textbf{n})=s_{11}(\textbf{m}\textbf{u})^{2}(\textbf{n}\textbf{u})^{2}+s_{22}(\textbf{m}\textbf{v})^{2}(\textbf{n}\textbf{v})^{2}\nonumber \\
+2s_{33}(\textbf{m}\textbf{u})(\textbf{m}\textbf{v})(\textbf{n}\textbf{u})(\textbf{n}\textbf{v})
+s_{12}\left[(\textbf{m}\textbf{u})^{2}(\textbf{n}\textbf{v})^{2}+(\textbf{m}\textbf{v})^{2}(\textbf{n}\textbf{u})^{2}\right]\nonumber \\
+\sqrt{2}s_{13}\left[(\textbf{m}\textbf{u})^{2}(\textbf{n}\textbf{u})(\textbf{n}\textbf{v})+(\textbf{n}\textbf{u})^{2}(\textbf{m}\textbf{u})(\textbf{m}\textbf{v})\right]\nonumber \\
+\sqrt{2}s_{23}\left[(\textbf{m}\textbf{v})^{2}(\textbf{n}\textbf{u})(\textbf{n}\textbf{v})+(\textbf{n}\textbf{v})^{2}(\textbf{m}\textbf{u})(\textbf{m}\textbf{v})\right],
	\label{nu-oblique}
\end{eqnarray} 
\begin{eqnarray}
\left[4G(\textbf{m},\textbf{n})\right]^{-1}=s_{11}(\textbf{m}\textbf{u})^{2}(\textbf{n}\textbf{u})^{2}+s_{22}(\textbf{m}\textbf{v})^{2}(\textbf{n}\textbf{v})^{2}\nonumber \\
+s_{33}\left[(\textbf{m}\textbf{u})(\textbf{n}\textbf{v})+ (\textbf{m}\textbf{v})(\textbf{n}\textbf{u})\right]^{2}/2
+2s_{12}(\textbf{m}\textbf{u})(\textbf{n}\textbf{u})(\textbf{m}\textbf{v})(\textbf{n}\textbf{v})\nonumber \\
+\sqrt{2}\left[s_{13}(\textbf{m}\textbf{u})(\textbf{n}\textbf{u})+s_{23}(\textbf{m}\textbf{v})(\textbf{n}\textbf{v})\right]\left[(\textbf{m}\textbf{u})(\textbf{n}\textbf{v})+(\textbf{m}\textbf{v})(\textbf{n}\textbf{u})\right].
	\label{shear-oblique}
\end{eqnarray}

It should be noticed that vectors of stretch $\textbf{n}$ and strain $\textbf{m}$ are two mutually perpendicular vectors of the unit length. In the reference frame related to the symmetry axes their components fulfill three conditions
\begin{equation}
	(\textbf{n}\textbf{u})^{2}+(\textbf{n}\textbf{v})^{2}=1,\, (\textbf{m}\textbf{u})^{2}+(\textbf{m}\textbf{v})^{2}=1,\, (\textbf{n}\textbf{u})(\textbf{m}\textbf{u})+(\textbf{n}\textbf{v})(\textbf{m}\textbf{v})=0.
	\label{eq:nm-cond}
\end{equation}
This means that only one component of $\textbf{n}$ and $\textbf{m}$ is independent. That implies, via (\ref{eq:nm-cond}), that $m_{1}=-n_{2}=-\sin\varphi$, and $m_{2}=n_{1}=\cos\varphi$. The explicit expressions for elastic material constants of 2D crystalline media and their consequences are studied in a companion paper \cite{cztas}.  
\section*{Appendix: the Christoffel tensor}
The Christoffel (sound propagation) tensor $\Gamma (\textbf{n})$ is defined as \cite{every}  
\begin{equation}
	\Gamma (\textbf{n})=\textbf{n}\cdot C\cdot\textbf{n}. \nonumber
\end{equation}

The Christoffel equation allows one to find the phase velocities $c(\textbf{n})$ and polarization vectors $\textbf{U}$ of a crystalline medium with the density $\rho$
\begin{equation}
	\left(\Gamma_{rs}-\rho c^{2}\delta_{rs}\right)U_{s}=0.
\end{equation}

With the help of Eq. (\ref{eq:parallell-tensor-expansion}) for oblique system, we obtain 
\begin{equation}
	\Gamma_{\rm{o}}(\mathbf{n})=\gamma_{uu}(\textbf{n})(\textbf{u}\otimes\textbf{u})+\gamma_{vv}(\textbf{n})(\textbf{v}\otimes\textbf{v})+ \gamma_{uv}(\textbf{n})(\textbf{u}\otimes\textbf{v}+\textbf{v}\otimes\textbf{u}),\nonumber
\end{equation}
where 
\begin{eqnarray}
\gamma_{uu}^{(\rm{o})}(\textbf{n})=c_{11}(\textbf{n}\textbf{u})^{2}+c_{33}(\textbf{n}\textbf{v})^{2}/2+\sqrt{2}c_{13}(\textbf{n}\textbf{u})(\textbf{n}\textbf{v}),\nonumber \\
\gamma_{vv}^{(\rm{o})}(\textbf{n})=c_{22}(\textbf{n}\textbf{v})^{2}+c_{33}(\textbf{n}\textbf{u})^{2}/2+\sqrt{2}c_{23}(\textbf{n}\textbf{u})(\textbf{n}\textbf{v}),\nonumber \\ \gamma_{uv}^{(\rm{o})}(\textbf{n})=\gamma_{vu}^{(\rm{o})}(\textbf{n})=\left(c_{12}+c_{33}/2\right)(\textbf{n}\textbf{u})(\textbf{n}\textbf{v})\nonumber \\+c_{13}(\textbf{n}\textbf{u})^{2}/\sqrt{2}+c_{23}(\textbf{n}\textbf{v})^{2}/\sqrt{2}. \nonumber
\end{eqnarray}

For rectangular, cubic and isotropic systems, $c_{13}=c_{23}=0$. For quadratic system $c_{11}=c_{22}$, and for isotropic media $c_{L}=c_{M}$. 

In CRA one obtains
\begin{eqnarray}
\gamma_{11}^{(\rm{o})}(\textbf{n})=C_{11}n_{1}^{2}+C_{66}n_{2}^{2}+C_{16}n_{1}n_{2},&\nonumber \\
\gamma_{22}^{(\rm{o})}(\textbf{n})=C_{22}n_{2}^{2}+C_{66}n_{1}^{2}+C_{26}n_{1}n_{2},&\nonumber \\ \gamma_{12}^{(\rm{o})}(\textbf{n})=\gamma_{21}^{(\rm{o})}(\textbf{n})=\left(C_{12}+C_{66}\right)n_{1}n_{2}
+C_{16}n_{1}^{2}+C_{26}n_{2}^{2}.& \nonumber
\end{eqnarray}
Taking into account that for rectangular and quadratic systems $C_{16}=C_{26}=0$, and additionally that for quadratic system $C_{11}=C_{22}$, one obtains Christoffel's matrix for these systems. If one takes $C_{66}=\left(C_{11}-C_{12}\right)/2$, then one obtains Christoffel's matrix for isotropic media. On the other hand, using Eq. (\ref{eq:A-decomp}) one obtains 
 
	\begin{equation*}
	\Gamma_{is}(\textbf{n}) =\left(
\begin{array}[c]{ll}
c_{J}n_{1}^{2}+c_{K} 		& c_{J}n_{1}n_{2}\\
c_{J}n_{1}n_{2} 				& c_{J}n_{2}^{2}+c_{K}
\end{array}
\right),
\end{equation*}
with $c_{J}$ and $c_{K}$ given by Eqs. (\ref{eq:C-S-exp-coeff}). Christoffel's tensors of isotropic 2D media obtained on the above two ways are identical. 

The propagation vector $\textbf{n}=\textbf{U}_{L}$ and the vector $\textbf{q}=\textbf{U}_{t}$ perpendicular to $\textbf{n}$ are the polarization vectors. The corresponding phase velocities are 
\begin{eqnarray}
	c_{l}=\sqrt{\left(C_{J}+C_{K}\right)/2\rho}=\sqrt{C_{11}/\rho},\nonumber\\
	c_{t}=\sqrt{C_{K}/2\rho}=\sqrt{\left(C_{11}-C_{12}\right)/2\rho}.\nonumber
\end{eqnarray}

The Christoffel tensor and phase velocities for quadratic system are discussed in a separate paper \cite{cztas}.

\end{document}